\documentclass[sigconf,authorversion,nonacm]{acmart}
\AtBeginDocument{%
  }

\usepackage{url}

\setcopyright{acmlicensed}
\copyrightyear{2018}
\acmYear{2018}
\acmDOI{XXXXXXX.XXXXXXX}
\acmConference[Conference acronym 'XX]{Make sure to enter the correct
  conference title from your rights confirmation emai}{June 03--05,
  2018}{Woodstock, NY}
\acmISBN{978-1-4503-XXXX-X/18/06}

\begin{document}

\title{An Annotated Reading of `The Singer of Tales' in the LLM Era}

\author{Kush R. Varshney}
\email{krvarshn@us.ibm.com}
\affiliation{%
  \institution{IBM Research -- Thomas J. Watson Research Center}
  \city{Yorktown Heights}
  \state{New York}
  \country{USA}
}


\begin{abstract}
The Parry-Lord oral-formulaic theory was a breakthrough in understanding how oral narrative poetry is learned, composed, and transmitted by illiterate bards. In this paper, we provide an annotated reading of the mechanism underlying this theory from the lens of large language models (LLMs) and generative artificial intelligence (AI). We point out the the similarities and differences between oral composition and LLM generation, and comment on the implications to society and AI policy.
\end{abstract}

\ccsdesc[500]{Computing methodologies~Philosophical/theoretical foundations of artificial intelligence}
\keywords{Artificial Intelligence, Mechanism, Oral Narrative Poetry}


\maketitle

\section{Introduction}
We have reached an era in which large language models (LLMs) based on the transformer architecture have become the de facto standard approach for many artificial intelligence (AI) tasks, especially generative AI. However, we still do not understand their basic mechanism with a simplicity that would allow five-year-olds and policymakers to understand \cite{Jensen2024}. LLMs have been compared to blurry JPEGs of the web \cite{hubert2024current}, mirrors of society \cite{Vallor2024}, magic 8 balls \cite{Heaven2024}, stochastic parrots \cite{bender2021dangers}, and disc jockeys \cite{lu2024ai}. The first two analogies are not mechanisms and the latter three leave out the desired level of precision and accuracy to be useful mechanisms. We ask: Is the right comparison of LLMs to the traditional oral narrative poet, the bard, the singer of tales \cite{benzon2023chatgpt}? 

Milman Parry and Alfred Lord disrupted Homeric scholarship, and by extension, the understanding of epic poetry the world over by proposing a new mechanism of language generation \cite{jensen2017challenge} they discovered by studying the living oral poetry tradition in Yugoslavia. Lord summed up their findings in the book \underline{The Singer of Tales}, published in 1960 \cite{Lord1960}. Known as \emph{oral-formulaic theory}, they found that narrative poetry is created orally by illiterate bards\footnote{Google denoted one of its early LLMs Bard, but we do not attempt to interpret the significance or connotation of this choice \cite{Barthes1972}.} at the time it is performed, based on formulas and themes. In this paper, we will go into the depths of the Parry-Lord theory and its explanatory power for LLMs by providing a reading of \underline{The Singer of Tales} annotated with observations about LLMs. In doing so, we will also comment on LLMs' larger impacts on society and public policy, including on aesthetics, authorship, copyright, use cases, cost, and guardrails.

The remainder of the paper alternates between verbatim passages selected from Lord (1960) in italic type, and our commentary in plain roman type preceded by a bullet marker. The section titles are chapter titles from \underline{The Singer of Tales}. Chapters 8--10 are omitted as they present detailed applications of the oral-formulaic theory to the Odyssey, Iliad, and medieval epics.

\section{Foreword}

\noindent\textit{  
We believe that the epic singers from the dawn of human consciousness have been a deeply significant group and have contributed abundantly to the spiritual and intellectual growth of man.
}

$\bullet$ We believe that LLMs are poised to have the same profound effect on human culture as the singers of tales. It is imperative to not let the moment pass us by without trying to understand what that effect might be. It may seem odd to look at an old tradition to learn about modern technology, but as they say, history does not repeat itself, but it often rhymes.

\vspace{8pt}\noindent\textit{
This book concentrates on only one aspect of the singers' art. Our immediate purpose is to comprehend the manner in which they compose, learn, and transmit their epics. It is a study in the processes of composition of oral narrative poetry.
}

$\bullet$ Similarly, this paper focuses on LLMs' mechanisms of learning and generating natural language. A \emph{mechanism} is a fundamental understanding of a phenomenon like plate tectonics as the mechanism for continental drift and other phenomena in earth science, the germ theory as the mechanism for the spread of diseases, and natural selection as the mechanism of evolution.\footnote{Parry is known as the Darwin of Homeric studies and oral tradition.} A mechanism must be explanatory, counterfactual, and constructive \cite{craver2013search}. A mechanism of this kind is quite high-level --- much higher-level than what is considered in the field of mechanistic interpretability at present \cite{bereska2024mechanistic}. A basic mechanism for LLMs may help policymakers reach decisions on AI use, safety, attribution, and copyright \cite{Jensen2024}.

\section{Chapter One, Introduction}

\noindent\textit{
in spite of the number of books about Homer and his poems, about epic poetry in general, and about specific epic traditions in various parts of the world, the student of epic still lacks a precise idea of the actual technique of \emph{poiesis} in its literal meaning.
}

$\bullet$ Poiesis is the process of creating something new artistically or intellectually. Generative AI also creates new artifacts. Although we full well know its process in terms of matrix-vector multiplications and model structures \cite{cho2024transformer}, we do not yet have much of a satisfactory answer on how higher-level LLM behaviors emerge from those lower-level operations and abstractions.

\vspace{8pt}\noindent\textit{
Stated briefly, oral epic song is narrative poetry composed in a manner evolved over many generations by singers of tales who did not know how to write; it consists of the building of metrical lines and half lines by means of formulas and formulaic expressions and of the building of songs by the use of themes.
}

$\bullet$ LLMs based on the transformer architecture are trained on enormous amounts of written content, which serves as their analog to lineages of singers. They generate natural language one token at a time through the attention mechanism: known simply as \emph{next-token prediction}. Do LLMs write? This is a question never asked in this way before, but we will attempt to answer it herein.

\vspace{8pt}\noindent\textit{
His manner of composition differs from that used by a writer in that the oral poet makes no conscious effort to break the traditional phrases and incidents; he is forced by the rapidity of composition in performance to use these traditional elements. To him they are not merely necessary, however; they are also right.
}

$\bullet$ LLMs also generate language in a single pass with low latency. The LLM's process, known as inference, is thus very much more like an oral poet than a writer. Empirical analysis shows that LLMs' output is characteristically different from human writing \cite{reinhart2024llms}. 

But is this necessary? Newer approaches are considering \emph{inference-time compute} that allows LLMs to select one among several possible outputs based on some criteria or even edit responses \cite{manvi2024adaptive}, thereby bringing them closer to writers. However, since inference-time compute follows the first pass of LLM generation as a post-processing step, it is important for us to focus our analysis on that first pass, which we can say is not writing but akin to oral composition. 

Despite alternative views \cite{porter2024moral}, most accounts do not endow LLMs an intrinsic sense of right and wrong. Thus it is not valid to ask whether an LLM believes its method of inference to be right. Even niche neural architecture search methods that use LLMs to determine AI structure and inference methods cannot report whether next-token prediction is \emph{right} \cite{nasir2024llmatic}. In a broader sense, there is an anthropomorphism risk in relating LLMs to bards \cite{ShneidermanM2023}. To avoid this risk when understanding the oral poet's methods of learning, composition, and transmission, we will treat the poet more like a machine than treating the LLM like a person. 

\vspace{8pt}\noindent\textit{
With oral poetry we are dealing with a particular and distinctive process in which oral learning, oral composition, and oral transmission almost merge; they seem to be different facets of the same process.
}

$\bullet$ The same is true of LLMs. We cannot understand their inference without also understanding their training.

\vspace{8pt}\noindent\textit{
Any terms, also, carrying implications derogatory to either oral narrative poetry or written poetry (as, for example, such terms as "authentic" and "artificial"; "primary" and "secondary") must be abandoned, for they represent an attitude that is neither scholarly nor critical. Both these forms are artistic expressions, each with its own legitimacy. We should not seek to judge but to understand.
}

$\bullet$ AI-created text has emerged as one more category beyond oral narrative text and written text. It has been described derogatorily as `AI slop.' But this is a misapplication of the aesthetics developed for written text. Even recent work to improve textual aesthetics of AI-generated content bases its methodology on the aesthetics of written text \cite{jiang2024textual}. In contrast, Gurdeep Pall, co-founder of Intelliflicks Studios, says, ``You have to look at [AI] like a new medium. You can’t paint the same thing with watercolors as you can with oil'' \cite{Gent2024}. We call for a new aesthetics for AI-created content.

\vspace{8pt}\noindent\textit{
It is a strange phenomenon in intellectual history as well as in scholarship that the great minds herein represented, minds which could formulate the most ingenious speculation, failed to realize that there might be some other way of composing a poem than that known to their own experience. They knew and spoke often of folk ballad and epic, they were aware of variants in these genres, yet they could see only two ways in which those variants could come into being: by lapse of memory or by wilful change. This seemed so obvious, so much an unquestioned basic assumption, that they never thought to investigate exactly how a traditional poetry operated. They always thought in terms of a fixed text or a fixed group of texts to which a poet \emph{did} something for a reason within his own artistic or intellectual self. They could not conceive of a poet composing a line in a certain way because of necessity or because of the demands of his traditional art.
}

$\bullet$ A mechanism cannot be superficial by describing a behavior without describing how the behavior comes about \cite{craver2013search}. Complicated equations involving deferents and epicycles developed under a geocentric model of planets describe the motion of planets alright, but do not describe how the system works. We seek a description of how LLMs work.

\section{Chapter Two, Singers: Performance and Training}

\noindent\textit{
In the case of a literary poem there is a gap in time between composition and reading or performance; in the case of the oral poem this gap does not exist, because composition and performance are two aspects of the same moment. }

$\bullet$ The LLM follows the same pattern as the oral poet. Its responses are consumed immediately in a prompt studio, assistant or copilot.

\vspace{8pt}\noindent\textit{
When we realize that the performance is a moment of creation for the singer, we cannot but be amazed at the circumstances under which he creates.
}

$\bullet$ There was wonder and awe surrounding ChatGPT when it was first released two years ago, even among AI practitioners and experts. However, that amazement has given way to disillusionment among many who view it instrumentally and are not finding as many productive uses for it as they imagined. 

\vspace{8pt}\noindent\textit{
He is not really a professional, but his audience does buy him drinks, and if he is good they will give him a little money for the entertainment he has given them.
}

$\bullet$ The original hope had been that LLMs would immediately remake numerous professions. But if we equate the LLM to the singer of tales, the generic LLM's role is not as a professional, but as an entertaining demonstration. The original ChatGPT and today's prevalent LLMs display a very generic helpfulness because they are not targeted to any use case. As we move toward smaller LLMs fine-tuned for specific use cases and agentic AI workflows, professionalism and sustainable businesses may emerge.

\vspace{8pt}\noindent\textit{
Whether the performance takes place at home, in the coffee house, in the courtyard, or in the halls of a noble, the essential element of the occasion of singing that influences the form of the poetry is the variability and instability of the audience.
}

$\bullet$ Tuning to the context, audience, and use case is imperative for effective LLMs.

\vspace{8pt}\noindent\textit{
If we are fully aware that the singer is composing as he sings, the most striking element in the performance itself is the speed with which he proceeds. It is not unusual for a Yugoslav bard to sing at the rate of from ten to twenty ten-syllable lines a minute. Since, as we shall see, he has not memorized his song, we must conclude either that he is a phenomenal virtuoso or that he has a special technique of composition outside our own field of experience.
}

$\bullet$ We are struck because the oral poet is human like us. We are not similarly struck by LLMs because of their supercomputer substrate. However, LLMs are not an inevitability just because of advanced computing technology. The technique is important.

\vspace{8pt}\noindent\textit{
the singer of fame will make a deeper impression on the tradition than will others of less repute.
}

$\bullet$ Some famous LLMs, like Meta's Llama, have made a deep impression as the base upon which hundreds of thousands of other LLMs have been fine-tuned \cite{foley2023matching}. Other famous models, like OpenAI's GPT-4, have made a deep impression by being the source of training data of other LLMs; many LLMs reply ``GPT-4'' if you prompt them ``Who are you?''

\vspace{8pt}\noindent\textit{
The oral singer in Yugoslavia is not marked by a class distinction; he is not an oral poet because he is a farmer or a shopkeeper or a bey. He can belong to the "folk," the merchant class, or the aristocracy.
}

$\bullet$ Although limited to males, the socioeconomic diversity of singers of tales is different from the situation in AI. Since developing LLMs requires immense resources, only extremely well-funded tech companies --- the aristocracy --- are able to do so. Subsequently, high-resourced perspectives are over-represented in LLMs \cite{choudhury2023generative}. This phenomenon, which comes from both pre-training and post-training (i.e.\ fine-tuning and alignment) \cite{qi2023fine}, is a serious problem that may be viewed as a kind of coloniality \cite{varshney2024decolonial}.

\vspace{8pt}\noindent\textit{
when writing is introduced and begins to be used for the same purpose as the oral narrative song, when it is employed for telling stories and is widespread enough to find an audience capable of reading, this audience seeks its entertainment and instruction in books rather than in the living songs of men, and the older art gradually disappears. The songs have died out in the cities not because life in a large community is an unfitting environment for them but because schools were first founded there and writing has been firmly rooted in the way of life of the city dwellers.
}

$\bullet$ Video killed the radio star just as writing killed the oral tradition because the newer technology began to occupy the same niche as the older technology. Creatives today worry that their work will be killed by LLMs. But are LLMs and their output really beginning to occupy the same niche as creative writers and complex writing? It does not seem so given that the top use cases for LLMs are business applications like question answering, text summarization, personalized marketing content generation, language translation, customer service support, market research analysis, document classification, and other non-creative endeavors. LLMs are assisting creative writers in inspiration, translation, and perspective, but not replacing them \cite{gero2023ai}. However, there are certainly other roles in which the aesthetics of literature are not needed and people will gradually disappear.

\vspace{8pt}\noindent\textit{
We can trace three distinct stages in his progress [learning the art]. During the first period he sits aside while others sing. He has decided that he wants to sing himself, or he may still be unaware of this decision and simply be very eager to hear the stories of his elders. Before he actually begins to sing, he is, consciously or unconsciously, laying the foundation.
}

$\bullet$ The first stage of learning by the singer of tales is equivalent to data collection and curation in LLMs. 

\vspace{8pt}\noindent\textit{
how does the oral poet meet the need of the requirements of rapid composition without the aid of writing and without memorizing a fixed form? His tradition comes to the rescue. Other singers have met the same need, and over many generations there have been developed many phrases which express in the several rhythmic patterns the ideas most common in the poetry. These are the formulas of which Parry wrote. In this second stage in his apprenticeship the young singer must learn enough of these formulas to sing a song. He learns them by repeated use of them in singing, by repeatedly facing the need to express the idea in song and by repeatedly satisfying that need, until the resulting formula which he has heard from others becomes a part of his poetic thought. He must have enough of these formulas to facilitate composition.
}

$\bullet$ The repeated use of formulas in singing by young singers is equivalent to pre-training LLMs by self-supervised learning: predicting text that fills masked sections of longer texts.

\vspace{8pt}\noindent\textit{
Learning in this second stage is a process of imitation, both in regard to playing the instrument and to learning the formulas and themes of the tradition. It may truthfully be said that the singer imitates the techniques of composition of his master or masters rather than particular songs. For that reason the singer is not very clear about the details of how he learned his art, and his explanations are frequently in very general terms. He will say that he was interested in the old songs, had a passion (\emph{merak}) for them, listened to singers, and then, "work, work, work" (\emph{goni, goni, goni}), and little by little he learned to sing. He had no definite program of study, of course, no sense of learning this or that formula or set of formulas. It is a process of imitation and of assimilation through listening and much practice on one's own.
}

$\bullet$ Pre-training LLMs is also \emph{goni, goni, goni}, requiring months of computations on thousands of GPUs! Similar to the singer, the details of LLM training dynamics are not very clear \cite{redman2024identifying}.

\vspace{8pt}\noindent\textit{
The second stage ends when the singer is competent to sing one song all the way through for a critical audience. There are probably other songs that he can sing partially, songs that are in process of being learned. He has arrived at a definite turning point when he can sit in front of an audience and finish a song to his own satisfaction and that of the audience. His job may or may not be a creditable one. 
}

$\bullet$ Some of the early language models like RoBERTa and GPT-2 are competent in simpler use cases like classification or entity resolution. Similarly, larger models have rudimentary capabilities early in their training. 

\vspace{8pt}\noindent\textit{
He has very likely not learned much about "ornamenting" a song to make it full and broad in its narrative style. That will depend somewhat on his model. If the singer from whom he has learned is one who uses much "ornamentation," he has probably picked up a certain amount of that ornamentation too.
}

$\bullet$ We can draw a congruence between ornamenting a song and making an LLM helpful. Helpfulness is usually introduced to LLMs using alignment methods after pre-training; however, similar to the bard's second stage of learning potentially including ornamentation, LLM training data may include some helpfulness data.

\vspace{8pt}\noindent\textit{
If however and this is important, he has not learned it from one singer in particular, and if the stories of that song differ in the various versions which he has heard, he may make a composite of them. He may, on the other hand, follow one of them for the most part, taking something from the others too. Either way is consistent with the traditional process. One can thus see that although this process should not be described as haphazard, which it is not, it does not fit our own conceptions of learning a fixed text of a fixed song.
}

$\bullet$ The oral poet is not a stochastic parrot just repeating his training as a fixed text. We should not oversimplify LLMs as stochastic parrots either \cite{bender2021dangers}.

\vspace{8pt}\noindent\textit{
His unlettered state saves him from becoming an automaton.
}

$\bullet$ Parrots and automatons represent the same concept --- something or someone that mechanically produces outputs by rote. The most interesting aspect of the Parry-Lord theory is the key role of illiteracy in learning and composition. Given all of the similarities already pointed out between oral narrative poets and LLMs, we must ask whether LLMs are also illiterate in some fashion.

\vspace{8pt}\noindent\textit{
Our proper understanding of these procedures is hindered by our lack of a suitable vocabulary for defining the steps of the process. The singers themselves cannot help us in this regard because they do not think in terms of form as we think of it; their descriptions are too vague, at least for academic preciseness. Man without writing thinks in terms of sound groups and not in words, and the two do not necessarily coincide. When asked what a word is, he will reply that he does not know, or he will give a sound group which may vary in length from what we call a word to an entire line of poetry, or even an entire song. The word for "word" means an "utterance." When the singer is pressed then to say what a line is, he, whose chief claim to fame is that he traffics in lines of poetry, will be entirely baffled by the question; or he will say that since he has been dictating and has seen his utterances being written down, he has discovered what a line is, although he did not know it as such before, because he had never gone to school.
}

$\bullet$ Opposite of the main flow of the paper thus far, AI provides us with a vocabulary to better understand the process of the singer of tales. The word or sound group of the bard may be viewed as a \emph{token} for the LLM: the smallest unit of processing that may be a part of a word or a whole word. Tokenization algorithms usually have fixed outputs, but newer methods may vary the notion of the token depending on uncertainty or information content with extensions of single tokens to longer phrases \cite{pagnoni2024byte}. Since LLMs traffic in tokens, which are then embedded in an abstract space, it is not surprising to see phenomena like the following. ChatGPT was famously prompted ``Hey, ChatGPT, how many R's are there in the word 'strawberry'?'' and responded in bafflement ``There are two R's in the word {'strawberry.'}'' \cite{fu2024large}. We can rightly say that the LLM is as illiterate as the singer of tales.

\vspace{8pt}\noindent\textit{
[In the third stage, w]hile the singer is adding to his repertory of songs, he is also improving the singing of the ones he already knows, since he is now capable of facing an audience that will listen to him, although possibly with a certain amount of patronizing because of his youth. Generally speaking, he is expanding his songs in the way I have indicated, that is, by ornamenting them.
}

$\bullet$ The third stage for the oral poet corresponds to the fine-tuning and alignment that LLMs undergo for following instructions and being helpful. This stage for the LLM is a combination of supervised learning and reinforcement learning. The audience is the source of the feedback data for the tuning.

\vspace{8pt}\noindent\textit{
Here, then, for the first time the audience begins to play a role in the poet's art. Up to this point the form of his song has depended on his illiteracy and on the need to compose rapidly in the traditional rhythmic pattern. The singers he has heard have given him the necessary traditional material to make it possible for him to sing, but the length of his songs and the degree to which he will ornament and expand them will depend on the demands of the audience. His audience is gradually changing from an attitude of condescension toward the youngster to one of accepting him as a singer.
}

$\bullet$ Initially while alignment of an LLM is ongoing, the audience is not actual users because they too would be condescending like the audience is to young singers. Outsourced data annotators provide the initial feedback instead. Later, once the LLM meets quality checks and is released or deployed, feedback continues to accrue from real users.

\vspace{8pt}\noindent\textit{
When he has a sufficient command of the formula technique to sing any song that he hears, and enough thematic material at hand to lengthen or shorten a song according to his own desires and to create a new song if he sees fit, then he is an accomplished singer and worthy of his art. There are, to be sure, some singers, not few in number, who never go beyond the third stage in learning, who never reach the point of mastery of the tradition, and who are always struggling for competence.
}

$\bullet$ Only LLMs with sufficient data collection and curation, pre-training, and post-training alignment are highly performant and can be used in advanced use cases.

\vspace{8pt}\noindent\textit{
The singer never stops in the process of accumulating, recombining, and remodeling formulas and themes, thus perfecting his singing and enriching his art. He proceeds in two directions: he moves toward refining what he already knows and toward learning new songs. The latter process has now become for him one of learning proper names and of knowing what themes make up the new song. The story is all that he needs; so in this stage he can hear a song once and repeat it immediately afterwards --- not word for word, of course --- but he can tell the same story again in his own words. Sometimes singers prefer to have a day or so to think the song over, to put it in order, and to practice it to themselves. Such singers are either less confident of their ability, or they may be greater perfectionists.
}

$\bullet$ Released LLMs are further improved for a given task by either in-context learning through prompting or by parameter-efficient or full fine-tuning. The singer who immediately retells a story is like an in-context learner and the one who takes a day to think it over is like LLMs that undergo fine-tuning. Just like with the bards, fine-tuning LLMs usually leads to better results at the expense of time and computation.

\vspace{8pt}\noindent\textit{
Zogić learned from Makić the song under discussion in his conversation, and both versions are published in Volume I of the Parry Collection (Nos. 24-25 and 29). Zogić did not learn it word for word and line for line, and yet the two songs are recognizable versions of the same story. They are not close enough, however, to be considered "exactly alike." Was Zogić lying to us? No, because he was singing the story as he conceived it as being "like" Makić's story, and to him "word for word and line for line" are simply an emphatic way of saying "like." As I have said, singers do not know what words and lines are. What is of importance here is not the fact of exactness or lack of exactness, but the constant emphasis by the singer on his role in the tradition. It is not the creative role that we have stressed for the purpose of clarifying a misunderstanding about oral style, but the role of conserver of the tradition, the role of the defender of the historic truth of what is being sung; for if the singer changes what he has heard in its essence, he falsifies truth. It is not the artist but the historian who speaks at this moment, although the singer's concept of the historian is that of a guardian of legend.
}

$\bullet$ In addition to helpfulness, a key tenet of LLM model providers is honesty, faithfulness or groundedness to source materials, and a lack of hallucination. In LLMs as well as in singers of tales, the goal of truth is in a semantic sense, not in the sense of precise wording. The poet as conserver of tradition, defender of truth, and guardian of legend is commensurate with the prevalent use cases for LLMs discussed earlier. Complex creative writing that invents new legends is a different use case. Such creativity has a precise mathematical tradeoff with faithfulness and other facets of AI safety \cite{varshney2023banal}.

\vspace{8pt}\noindent\textit{
And the picture that emerges is not really one of conflict between preserver of tradition and creative artist; it is rather one of the preservation of tradition by the constant re-creation of it. The ideal is a true story well and truly retold.
}

$\bullet$ It is important to note that unique word choices and fluent turns of phrase are not what is meant by creativity in an LLM creativity-safety tradeoff analysis like \cite{varshney2023banal}. What is meant is creativity in the semantics --- something that may be measured using \emph{semantic entropy} \cite{farquhar2024detecting}. Therefore, ignoring differences in phrasing that have equivalent meaning, LLMs for top use cases and singers of tales operate in a common regime that can achieve both language fluency and low hallucination. LLMs, however, may be operated in more hallucinatory regimes as well. Either inadvertently or purposefully, LLMs may create misinformation that threatens fragile information ecosystems \cite{Shaily2025}. 

\section{Chapter Three, The Formula}

\noindent\textit{
There came a time in Homeric scholarship when it was not sufficient to speak of the "repetitions" in Homer, of the "stock epithets," of the "epic cliches" and "stereotyped phrases." Such terms were either too vague or too restricted. Precision was needed, and the work of Milman Parry was the culmination of that need. The result was a definition of the "formula" as "a group of words which is regularly employed under the same metrical conditions to express a given essential idea."
}

$\bullet$ Parry-Lord formulas are heuristic solutions to constrained optimization problems that must be solved in real-time. The constraints are preserving the essential meaning, having fluency of language, and fitting the metrical conditions. LLMs are rarely prompted to meet metrical constraints, but often have to meet other style or harmlessness constraints. Decoding, the part of LLM inference that produces actual response tokens from probabilities, is often heuristic and constrained as well. Greedy decoding produces many repetitions of the kind seen in Homer; beam search decoding does too, albeit less than greedy decoding. More elaborate decoding methods start encroaching on inference-time compute discussed earlier and cannot be done in real-time (and thereby become writing instead of oral composition). Thus, formulas are equally a part of LLMs as they are of the oral tradition. Moreover, they suit the needs of common LLM use cases like summarization and question answering that do not require long repetition-free responses.

\vspace{8pt}\noindent\textit{
The stress in Parry's definition on the metrical conditions of the formula led to the realization that the repeated phrases were useful not, as some have supposed, merely to the audience if at all, but also and even more to the singer in the rapid composition of his tale. And by this almost revolutionary idea the camera's eye was shifted to the singer as a composer and to his problems as such.
}

$\bullet$ The discovery of the oral-formulaic mechanism highlighted composition in performance under constraints and time pressure as the key to understanding why the oral songs are as they are. Decoding has not received as much attention as data curation, pre-training, and alignment in the world of LLMs. But by understanding the need for fitting constraints and for heuristic methods due to time/computation pressures, we may start to shift the AI world's camera's eye to decoding in single-pass inference.

\vspace{8pt}\noindent\textit{
If the singer is in the Yugoslav tradition, he obtains a sense of ten syllables followed by a syntactic pause, although he never counts out ten syllables, and if asked, might not be able to tell how many syllables there are between pauses. In the same way he absorbs into his own experience a feeling for the tendency toward the distribution of accented and unaccented syllables and their very subtle variations caused by the play of tonic accent, vowel length, and melodic line. These "restrictive" elements he comes to know from much listening to the songs about him and from being engrossed in their imaginative world.
}

$\bullet$ Although the constraints are different --- metrical ones for bards and style or harmlessness constraints for LLMs --- they are not explicit rules in either case. They are learned from examples or probability distributions and may be induced during decoding \cite{ko2024large,gu2024chared}. Thus even to this level of detail, LLMs have a mechanism quite similar to the oral-formulaic mechanism. 

\vspace{8pt}\noindent\textit{
The most stable formulas will be those for the most common ideas of the poetry. They will express the names of the actors, the main actions, time, and place.
}

$\bullet$ In the LLM hallucination analysis of Kalai and Vempala, such names and places are factoids. The rate of monofacts: uncommon factoids that appear rarely or only once in training provide a way to estimate the rate of hallucination because monofacts are almost indistinguishable from completely made up factoids (hence appearing only that one time) \cite{kalai2024calibrated}. 

Given that singers of tales are guardians of their tradition and do not wish for hallucinations to enter their songs, it makes sense for them to stick to common factoids because doing so serves to keep the hallucination rate down. In the LLM case, this indicates that it behooves users to stick to common queries if they are looking for low hallucination rates. It is often advised that users only ask questions of LLMs that they already know the answers to. The benefit of the LLM response is in its elaboration. Similarly, audiences of oral narrative poetry usually already know the story, but listen for the bard's elaboration.

\vspace{8pt}\noindent\textit{
[Formulas'] usefulness can be illustrated by indicating the many words that can be substituted for the key word in such formulas. For example, in the Prilip formulas above, any name of a city with a dative of three syllables can be used instead of Prilip: \emph{u Stambolu, u Travniku, u Kladuši}. Instead of \emph{a u kuli}, "in the tower," one can say \emph{a u dvoru}, "in the castle," or \emph{a u kući}, "in the house."
}

$\bullet$ LLMs are able to come up with such substitutions as well. This ability has been found useful in text perturbation-based source attribution methods that try to figure out key words in prompts or context that lead to given tokens in the response \cite{paes2024multi}.

\vspace{8pt}\noindent\textit{
Although it may seem that the more important part of the singer's training is the learning of formulas from other singers, I believe that the really significant element in the process is rather the setting up of various patterns that make adjustment of phrase and creation of phrases by analogy possible. This will be the whole basis of his art. Were he merely to learn the phrases and lines from his predecessors, acquiring thus a stock of them, which he would then shuffle about and mechanically put together in juxtaposition as inviolable, fixed units, he would, I am convinced, never become a singer.
}

$\bullet$ The emphasis on decoding in LLMs that emerges from an oral-formulaic view of inference is not all we need to explain the process of composition by LLMs. A critical piece of the transformer architecture underpinning LLMs is the attention mechanism \cite{vaswani2017attention}, which is precisely the mechanism that sets up the ``the various patterns that make adjustment of phrase and creation of phrases by analogy possible.'' Without it, LLMs would just be n-gram machines, not awe-inspiring generative AI. Decoding works in conjunction with the attention mechanism just like the analogous mechanisms of the oral narrative poet work together.

\vspace{8pt}\noindent\textit{
Under the pressure of rapid composition in performance, the singer of tales, it is to be expected, makes occasional errors in the construction of his lines. His text line may be a syllable too long or a syllable too short. This does not trouble him in performance, and his audience scarcely notices these lines, since they have an understanding of the singer's art and recognize these slight variations as perfectly normal aberrations.
}

$\bullet$ LLMs are not perfect in respecting their constraints either. For example, some harmful behaviors slip through into responses. This is why deployed LLMs are often paired with a second guardian model that checks the responses of the LLM \cite{zeng2024shieldgemma,padhi2024granite}. However, any action that may follow from a guardian model would constitute a form of editing or revision, and thus be the equivalent of writing rather than oral composition.

\vspace{8pt}\noindent\textit{
by necessity, because he does not remember all the phrases which he needs, he is forced at the moment of his private performances to form phrases on the basis of the patterns. Since they follow the traditional patterns, they are indistinguishable from the other phrases that he has remembered, and may unconsciously be actually identical with them. To him the first matter of importance is certainly not the source of the phrase but the phrase itself at the critical time.
}

$\bullet$ Although there is some amount of memorization of factoids in LLMs, they are shown to largely generalize rather than memorize overall \cite{wang2024generalization}. Because of this generalization, there is often no clear source attribution to response tokens. (We may try to conduct a post hoc attribution as mentioned earlier \cite{paes2024multi}.) As facts cannot be copyrighted and phrases from patterns are not original expressions of ideas, neither the singer's poem nor LLM responses should be considered subject to copyright. 

\vspace{8pt}\noindent\textit{
In order to avoid any misunderstanding, we must hasten to assert that in speaking of "creating" phrases in performance we do not intend to convey the idea that the singer \emph{seeks originality} or fineness of expression. He seeks expression of the idea under stress of performance. Expression is his business, not originality, which, indeed, is a concept quite foreign to him and one that he would avoid, if he understood it. To say that the \emph{opportunity} for originality and for finding the "poetically" fine phrase exists does \emph{not} mean that the \emph{desire} for originality also exists. There are periods and styles in which originality is \emph{not} at a premium. If the singer knows a ready-made phrase and thinks of it, he uses it without hesitation, but he has, as we have seen, a method of making phrases when he either does not know one or cannot remember one. 
}

$\bullet$ The singer of tales would not seek copyright, even if he could. Oral narrative poetry is simply different from literature in this respect; we submit that LLM responses are different from writing in the same way. LLMs are even poor at detecting originality \cite{huang2025large}. 

Emerging legal precedence contends that content generated solely by LLMs cannot be copyrighted, but the reason given is that only works created by humans can be copyrighted. We argue that human-involvement should not be the distinguishing characteristic, but whether originality is a part of the generation mechanism. Originality is not at a premium in the period and style of current LLM use cases: we are not asking LLMs to show originality in generating contracts or answering customer-support questions.

\vspace{8pt}\noindent\textit{
Such a living art, so closely united to individual experience, cannot help but leave its peculiar stamp upon the songs and their texts. Because of this mark left upon them we can with a high degree of certainty determine whether any text that is before us was formed by a traditional bard in the crucible of oral composition.
}

$\bullet$ Although sometimes maligned, detecting the difference between human-written text and LLM-generated text is possible to high accuracy \cite{hu2023radar}.

\vspace{8pt}\noindent\textit{
Although the formulas which any singer has in his repertory could be found in the repertories of other singers, it would be a mistake to conclude that all the formulas in the tradition are known to all the singers. There is no "check-list" or "handbook" of formulas that all singers follow.
}

$\bullet$ Methods for determining which particular LLM produced a given response similarly rely on differences in formulas (diversity in lexicon and parts of speech) \cite{kumarage2023neural}.

\vspace{8pt}\noindent\textit{
The syntactic and rhythmic parallelism of lines three and four modulates into a pattern of syntactic and rhythmic opposition in lines four and five, six and seven, eight and nine, at the same time that syntactic parallelism is kept between five and six, seven and eight. Had Ugljanin been a literate poet who sat down with pen in hand to devise these lines with their inner balances and syncopations, he could not have done better. One can even fancy the overliterate "interpreter of literature," innocent of Salih's ignorance of such matters, extolling the syncopation as the artful intent of the poet to indicate the zigzag search of the messenger for Alija!
}

$\bullet$ Literary criticism was progressing around the time that Lord published the \underline{Singer of Tales}. For example, Barthes posited that it is not the author's intentions that explain the meaning of a text \cite{barthes2016death}. The interpretation is fully in the hands of the audience. But in the Parry-Lord theory, oral poets are not literary authors at all and LLMs are not typically thought to have intentions. But can LLMs have any notion of authorship? We will return to this question later.

\vspace{8pt}\noindent\textit{
It is certainly possible that a formula that entered the poetry because its acoustic patterns emphasized by repetition a potent word or idea was kept after the peculiar potency which it symbolized and which one might say it even was intended to make effective was lost --- kept because the fragrance of its past importance still clung vaguely to it and kept also because it was now useful in composition. It is \emph{then} that the repeated phrases, hitherto a driving force in the direction of accomplishment of those blessings to be conferred by the story in song, began to lose their precision through frequent use. Meaning in them became vestigial, connotative rather than denotative. From the point of view of usefulness in composition, the formula means its essential idea; that is to say, a noun-epithet formula has the essential idea of its noun. The "drunken tavern" means "tavern." But this is only from the point of view of the singer composing, of the craftsman in lines.
}

$\bullet$ In the same time period as Lord and Barthes, L\'evi-Strauss proposed the structural study of myth with a focus on oral traditions \cite{levi1955structural}. He claimed that there may originally have been something important about, e.g. a coyote in Native American myth, but later it simply became a slot in a structure that could be filled with other tricksters. (Moreover, the structure is common across cultures so the trickster slot is filled by  Anansi the spider in West Africa, the sage N\=arada in India, and the god Loki among the Norse.) In this view, a noun-epithet formula fills a structural slot just as well as its noun. 

From the LLM perspective, if LLM-generated content that satisfied a constraint starts being used as training data for next-generation LLMs that do not have the same constraints, we may see model collapse yielding repetitiveness of tokens for no good reason. Separately and amplified by model collapse, such structures may lead LLMs to have unwanted biases as slots are filled with historical artifacts that perpetuate stereotypes and discrimination. We do not believe that bias in LLMs has previously been hypothesized to come from such a mechanism.

Moreover, in L\'evi-Strauss' conception, the structures of myths may be transformed mathematically within algebraic groups \cite{santucci2020discrete}. Experiments show that LLMs transform responses in a similar way upon prompting \cite{benzon2023chatgpt}. This also suggests using transformations within algebraic groups for mitigating biases in LLM responses, such as the equi-tuning method \cite{basu2023equi}.

\vspace{8pt}\noindent\textit{
The poet was sorcerer and seer before he became "artist." His structures were not abstract art, or art for its own sake. The roots of oral traditional narrative are not artistic but religious in the broadest sense.
}

$\bullet$ As we know, appropriate LLMs have an instrumental purpose in numerous use cases --- they are not just entertainment. Like oral narrative poetry, perhaps their broad instrumental use should be thought of as a structure to carry logical arguments forward. 

\section{Chapter Four, The Theme}

\noindent\textit{
Following Parry, I have called the groups of ideas regularly used in telling a tale in the formulaic style of traditional song the "themes" of the poetry.
}

$\bullet$ Oral-formulaic theory contains the higher-level construct of the theme. The analogous nomenclature fora group of ideas in LLMs is the \emph{concept}.

\vspace{8pt}\noindent\textit{
Incidents of this sort occur in song after song, and from much hearing the pattern of the theme becomes familiar to the youthful "bard even before he begins to sing. He listens countless times to the gathering of an army or of a large number of wedding guests (the two are often synonymous). He hears how the chieftain writes letters to other chiefs; he comes to know the names of these leaders of the past and of the places where they dwelt; he knows what preparations are made to receive the assembling host, and how each contingent arrives, what its heroes are wearing and what horses they are riding and in what order they appear. All this and much more is impressed upon him as he sits and is enthralled by his elders' singing of tales.
}

$\bullet$ Large concept models (LCMs) are a fairly new proposal in AI that use the transformer architecture on a second level of abstraction above tokens: concepts \cite{lcm2024large}. The initial work treats full sentences as concepts and embeds them into an existing sentence embedding space. In so doing, the concepts are also represented numerically agnostic to the underlying human language. The epic themes mentioned above by Lord recur in oral narrative poetry from different cultures, and like the LCM embedding, may be considered language-agnostic. Even emerging cultures like tech reuse the same themes: ``Some tech companies have become as powerful as empires and their leaders like emperors, as so often before, in search of an epic to consecrate their great deeds'' \cite{Reade2025}. We conjecture that an understanding of the oral-formulaic theory mechanism may have inspired LCMs earlier than their proposal in late 2024. 

\vspace{8pt}\noindent\textit{
Although he thinks of the theme as a unit, it can be broken down into smaller parts: the receipt of the letter, the summoning of the council, and so forth. Yet these are subsidiary to the larger theme. They will be useful perhaps in other contexts later on, but the singer learns them first for use in the specific council of the specific song, with the appropriate names of people and places and their characteristics. The names are attached in minor themes of calling the council, introducing speeches, in question and answer. All this the learner thinks through before he can be satisfied with his singing and before he can move on to the next larger theme.
}

$\bullet$ The training of LCMs follows the training of singers in themes, just as the training of LLMs follows the training of singers in formulas. 

\vspace{8pt}\noindent\textit{
With years of experience the singer becomes an active listener to the songs of others. The really talented oral poet combines listening and learning in one process. The listening is then dynamic and can be said to constitute in itself the first rehearsal of the new song. Singers who can do this are, however, rare. Many may boast, but their boast is a heroic one and belongs to the hyperboles of epic poetry. That it is possible I am sure; for I have seen and heard this marvel accomplished.
}

$\bullet$ Recent work shows that LLMs adapt their internal representations of concepts based on context, overriding the semantics learned during pre-training \cite{park2024iclr}. This in-context adaptation of LLMs is thus similar to the best among singers of tales.

\vspace{8pt}\noindent\textit{
In building a large theme the poet has a plan of it in his mind beyond the bare necessities of narrative. There are elements of order and balance within themes. The description of an assembly, for example, follows a pattern proceeding from the head of the assembly and his immediate retinue through a descending hierarchy of nobles to the cupbearer, who is the youngest in the assembly and hence waits upon his elders, but ending with the main hero of the story. This progression aids the singer by giving him a definite method of presentation.
}

$\bullet$ Planning is an increasingly sought out capability of LLMs with the emergence of agentic AI. LCMs are in fact being explored as planners \cite{lcm2024large}. This research direction in AI is not surprising given what we know about the singer of tales and the plans he is able to put together with themes.

\vspace{8pt}\noindent\textit{
The singer's mind is orderly.
}

$\bullet$ LLMs and LCMs are orderly too. We have just not had the right mechanistic understanding to describe that orderliness.

\vspace{8pt}\noindent\textit{
A singer ordinarily has one basic form for such a minor theme; it is flexible and within limits adaptable to special circumstances. But when such circumstances are absent, the singer makes no attempt to alter its general pattern.
}

$\bullet$ Under the oral-formulaic theory and real-world observations, bards use common minor themes even if they do not precisely fit the context. This is no different from higher-level conceptual hallucinations by LLMs.

\vspace{8pt}\noindent\textit{
In a traditional poem, therefore, there is a pull in two directions: one is toward the song being sung and the other is toward the previous uses of the same theme. The result is that characteristic of oral poetry which literary scholars have found hardest to understand and to accept, namely, an occasional inconsistency, the famous nod of a Homer.
}

$\bullet$ This pull to reuse themes due to the needs of the singer of tales seems to be a phenomenon that would arise in heuristic and constrained decoding of LCMs in the same way that token repetition arises in the decoding of LLMs. As the research on LCMs progresses, attention to decoding will be required.

\vspace{8pt}\noindent\textit{
there is a common stock of themes which we can conveniently label. But our neatly categorizing minds work differently from the singer's. To him the formulas and themes are always used in association one with another; they are always part of a song. 
}

$\bullet$ The interactions among formulas and themes in the bard's composition process point out the shortcomings of the magic 8 ball analogy of LLMs. Every shake of the magic 8 ball is independent, but not the next token coming out of an LLM. Moreover, LLM developers are increasingly attempting long-range attention mechanisms and long contexts to permit even more associations.

\section{Chapter Five, Songs and the Song}

\noindent\textit{
In some respects the larger themes and the song are alike. Their outward form and their specific content are ever changing. Yet there is a basic idea or combination of ideas that is fairly stable. We can say, then, that a song is the story about a given hero, but its expressed forms are multiple, and each of these expressed forms or tellings of the story is itself a separate song, in its own right, authentic and valid as a song unto itself.
}

$\bullet$ Despite having a different mechanism, the oral narrative story is like conceptual art that anyone may put together by following a set of instructions.\footnote{An example is the artwork "Untitled" (L.A.) by Felix Gonzalez-Torres, which is installed in museums as a pile of candies. Patrons are encouraged to take a candy and eat it. The piece has the instructions: Green candies individually wrapped in cellophane, endless supply, overall dimensions vary with installation, original weight: 50 lb (22.7 kg).} The art is in the concept, the idea. It challenges those of us accustomed to paintings, sculpture, and literature that have a fixed form and unambiguously-known constructor. In the LLM world, \emph{prompts} capture the concept or instructions of a human creator. A prompt leads to different responses from different LLMs, or even different inference runs of the same LLM due to randomness. For the conceptual artist and the singer of tales, each of those instantiations is the story.

\vspace{8pt}\noindent\textit{
Our real difficulty arises from the fact that, unlike the oral poet, we are not accustomed to thinking in terms of fluidity. We find it difficult to grasp something that is multiform. It seems to us necessary to construct an ideal text or to seek an original, and we remain dissatisfied with an ever-changing phenomenon. I believe that once we know the facts of oral composition we must cease trying to find an original of any traditional song. From one point of view each performance is an original. From another point of view it is impossible to retrace the work of generations of singers to that moment when some singer first sang a particular song.
}

$\bullet$ Why do we care about originals and authorship? As a counterpoint to Barthes, Foucault did not completely discount the author, but asked critical questions about the author function \cite{Foucault1969}. He argued that the author function is a social construct that comes about because society desires an attribution, especially so that the author is subject to punishment. As such, the author is more central in literature than scientific writing. He said that the author function is not universal, but is shaped by the context. 

If we extend this thinking to LLMs, with varying levels of human involvement, then perhaps there are more fine-grained ways of attributing authorship than simply a binary indicator. In fact, researchers have recently proposed AI attribution statements that detail the LLM used in co-creation, ``the types of contributions it made, the proportion of work it created or modified, and the initiative it took in helping produce the work'' \cite{He2025}.

\vspace{8pt}\noindent\textit{
The truth of the matter is that our concept of "the original," of "the song," simply makes no sense in oral tradition. To us it seems so basic, so logical, since we are brought up in a society in which writing has fixed the norm of a stable first creation in art, that we feel there must be an "original" for everything. The first singing in oral tradition does not coincide with this concept of the "original." We might as well be prepared to face the fact that we are in a different world of thought, the patterns of which do not always fit our cherished terms. In oral tradition the idea of an original is illogical.
}

$\bullet$ Looking back at the oral tradition and forward to co-creation with LLMs, we must change our norms because they are both in a different world of thought.

\vspace{8pt}\noindent\textit{
To the superficial observer, changes in oral tradition may seem chaotic and arbitrary. In reality this is not so. It cannot be said that "anything goes." Nor are these changes due in the ordinary sense to failure of memory of a fixed text, first, of course, because there is no fixed text, second, because there is no concept among singers of memorization as we know it, and third, because at a number of points in any song there are forces leading in several directions, any one of which the singer may take. If his experience of the particular song is weak, either as a whole or at any part, the force in a direction divergent from the one he has heard may be the strongest.
}

$\bullet$ As we have discussed, next token prediction in LLMs is the culmination of many forces ranging from data curation practices, to attention mechanisms, to supervised fine-tuning, to constrained decoding. We can mechanically follow probability distributions through this maze of steps \cite{cho2024transformer}, but the simplified view available to us through analogy with the singer of tales provides us a meaningful understanding of the process.

\vspace{8pt}\noindent\textit{
The fact that the same song occurs attached to different heroes would seem to indicate that the story is more important than the historical hero to which it is attached. There is a close relationship between hero and tale, but with some tales at least the type of hero is more significant than the specific hero. It is convenient to group songs according to their story content, or thematic configurations, because songs seem to continue in spite of the particular historical hero; they are not connected irrevocably to any single hero.
}

$\bullet$ The type of hero rather than the specific hero illustrates the theory of slot filling into the structure of myth proposed by L\'evi-Strauss mentioned earlier, and as discussed then, could be the source of hallucinations and biases in the LLM context.

\vspace{8pt}\noindent\textit{
When we look back over these examples of transmission, we are, I believe, struck by the conservativeness of the tradition. The basic story is carefully preserved. Moreover, the changes fall into certain clear categories, of which the following emerge: (1) saying the same thing in fewer or more lines, because of singers' methods of line composition and of linking lines together, (2) expansion of ornamentation, adding of details of description (that may not be without significance), (3) changes of order in a sequence (this may arise from a different sense of balance on the part of the learner, or even from what might be called a chiastic arrangement where one singer reverses the order given by the other), (4) addition of material not in a given text of the teacher, but found in texts of other singers in the district, (5) omission of material, and (6) substitution of one theme for another, in a story configuration held together by inner tensions.
}

$\bullet$ This summary of the oral-formulaic theory applies nearly as well to oral narrative poets as it does to LLMs with a single pass of inference, and provides us a different lens with which to understand the mechanism of LLMs.

\section{Chapter Six, Writing and Oral Tradition}

\noindent\textit{
The art of narrative song was perfected, and I use the word advisedly, long before the advent of writing. It had no need of stylus or brush to become a complete artistic and literary medium. Even its geniuses were not straining their bonds, longing to be freed from its captivity, eager for the liberation by writing. When writing was introduced, epic singers, again even the most brilliant among them, did not realize its "possibilities" and did not rush to avail themselves of it. 
}

$\bullet$ With capable LLMs emerging in the last couple of years, we now have a third paradigm of composition through LLMs that extends beyond oral and written composition (although oral composition is, for all intents and purposes, not a living tradition any more). Some tout potential productivity gains through generative AI with LLMs, but new technologies usually show a lag of years or even decades before affecting productivity growth \cite{brynjolfsson1993productivity}. Simpler use cases addressing non-writers in the business world are seeing adoption and operationalization of LLMs, but the majority of creative writers are not longing to be freed from the bondage of the writing process \cite{gero2023ai}.

\vspace{8pt}\noindent\textit{
It is vastly important that we do not make the unthinking mistake of believing that the process of dictation frees the singer to manipulate words in accordance with an entirely new system of poetics. Clearly he has time to plan his line in advance, but this is more of a hindrance than a help to a singer who is accustomed to rapid-fire association and composition. Opportunity does not make the singer into an e.e.cummings! not even if he is already a Homer! 
}

$\bullet$ LLMs are different from oral poets because highly performant LLMs \emph{can} be improved along several capability dimensions through memory-enhanced architectures \cite{das2024larimar}, inference-time compute \cite{manvi2024adaptive}, and metacognition more generally \cite{johnson2024imagining} that are the LLM equivalent of giving the bard reading and writing. However, such capability improvement has been shown in fact retrieval, hallucination reduction, mathematics, and reasoning rather than in quality of expression. 

\vspace{8pt}\noindent\textit{
From the point of view of verse-making, dictation carries no great advantage to the singer, but from that of song-making it may be instrumental in producing the finest and longest of songs. For it extends almost indefinitely the time limit of performance. And with a little urging, under the stimulus of great accomplishment for a worthy audience, the singer of talent will apply every resource of his craft to adorn and enrich his song. The important element is that of time; there is nothing in the dictating process itself that brings this richness to bear. The collector who tells a singer that he can sing his song from day to day taking as many days, as much time, as he wants, can elicit the same results in sung performance, as we saw in the case of Avdo Međedović's songs in the last chapter. It should be stressed also that the additional time is of use only to the exceptional singer of great talent in a tradition rich in traditional themes and songs.
}

$\bullet$ When the singer is in the special situation of dictating a story to a collector, the `use case' warrants extra time and resources. In LLMs, inference-time compute or memory-enhanced architectures may increase the time and computation cost beyond single pass inference. Another increase in time and computation cost is from back-and-forth multi-turn sessions in which the user prompts the LLM to revise and rewrite its responses. The use case and form of interaction with the LLM dictate whether taking extra time and incurring extra cost is warranted. Typical business use cases do not warrant the extra cost, but special circumstances in which the LLM is being asked to truly \emph{write} (or solve a difficult mathematics problem, etc.) may warrant it.

\vspace{8pt}\noindent\textit{
It is worthy of emphasis that the question we have asked ourselves is whether there can be such a thing as a transitional text; not a period of transition between oral and written style, or between illiteracy and literacy, but a text, product of the creative brain of a single individual. When this emphasis is clear, it becomes possible to turn the question into whether there can be a single individual who in composing an epic would think now in one way and now in another, or, perhaps, in a manner that is a combination of two techniques. I believe that the answer must be in the negative, because the two techniques are, I submit, contradictory and mutually exclusive.
}

$\bullet$ The oral tradition died out with the introduction of writing, but \emph{post-literate} traditions emerged that contain some aspects of oral composition within literate societies. As reported by Pihel, ``The term `post-literate,' then, acknowledges the historical progression from orality to literacy to post-literacy rather than a circular development back to orality; it implies that post-literate poetry both incorporates and exceeds literate poetry'' \cite{Pihel1996}. As the Parry-Lord theory predicts, they are not transitional, but a different third paradigm. One example is freestyling hip-hop rap \cite{Pihel1996,banks2010homer}.  In this hip-hop case, rhyme is a key constraint that is a product of writing. We view LLMs as another post-literate composition paradigm. 

\vspace{8pt}\noindent\textit{
Writing as a new medium will mean that the former singer will have a different audience, one that can read. Psychologically, he may at first be addressing himself still for some time to the audience of listeners to whom he has always been accustomed. But the new reading public, though it will be small at first, will undoubtedly have different tastes developing from those of the traditional nonliterate audience. They will demand new themes, or new twists to old themes.
}

$\bullet$ What will the audience of the LLM-composed medium have a taste for? Only time will tell.

\vspace{8pt}\noindent\textit{
The fact of writing does not inevitably involve a tradition of written literature
}

$\bullet$ Similarly, the existence of LLMs, including with memory and inference-time compute, does not necessarily imply that a tradition of AI-generated literature will emerge. As the field progresses and we continue to ask what LLMs are good for, our answer may end up being only the simpler non-literary use cases.

\vspace{8pt}\noindent\textit{
Oral tradition did not become transferred or transmuted
into a literary tradition of epic, but was only moved further and further into the background, literally into the back country, until it disappeared.
}

$\bullet$ The singer of tales disappeared. Will the creative writer now disappear? Or will post-literate co-creativity prosper alongside the tradition of writing?

\section{Chapter Seven, Homer}

\noindent\textit{
Each theme, small or large --- one might even say, each formula --- has around it an aura of meaning which has been put there by all the contexts in which it has occurred in the past. It is the meaning that has been given it by the tradition in its creativeness. To any given poet at any given time, this meaning involves all the occasions on which he has used the theme, especially those contexts in which he uses it most frequently; it involves also all the occasions on which he has heard it used by others, particularly by those singers whom he first heard in his youth, or by great singers later by whom he was impressed. To the audience the meaning of the theme involves its own experience of it as well. The communication of this suprameaning is possible because of the community of experience of poet and audience.
}

$\bullet$ The relationship between the poet and the audience has a depth of meaning that, risking anthropomorphism, we are starting to now observe between LLM and user. The relationship is deepened if there is a dyadic alignment or a mutual theory of mind between the user and the LLM \cite{varshney2025scopes}. This entails a level of explainability and understanding in both directions that is only possible if the lineage, provenance, and attribution of the sources of tokens and concepts is available. Earlier, we noted that in the oral-formulaic theory, the singer or the LLM loses track of the source in the moment of composition, but this metadata or meaning is not all lost and may be available in self-reflection. It is a critical element of the metacognitive parts of the tradition. 

\section{In Conclusion}

\noindent\textit{
Yet after all that has been said about oral composition as a technique of line and song construction, it seems that the term of greater significance is traditional. Oral tells us "how," but traditional tells us "what," and even more, "of what kind" and "of what force." When we know how a song is built, we know that its building blocks must be of great age. For it is of the necessary nature of tradition that it seek and maintain stability, that it preserve itself. And this tenacity springs neither from perverseness, nor from an abstract principle of absolute art, but from a desperately compelling conviction that what the tradition is preserving is the very means of attaining life and happiness.
}

$\bullet$ Famous LLMs are produced by tech companies that are the beacon of capitalism. Barthes, who we discussed earlier, saw capitalism as a system that perpetuates its dominance by creating epics and myths \cite{Barthes1972}. Similar to the self-preservation of the oral tradition, LLMs may very well end up serving as a form of self-preservation for tech by perpetuating its myths. 

Furthermore, Barthes saw the author as ``the epitome and culmination of capitalist ideology'' \cite{barthes2016death}. His view was shaped by authors as owners of their compositions that they could monetize through copyright and royalties. However as we have discussed throughout this reading, LLMs, like singers of tales, are not authors; in their basic form, they do not even write. Given that LLMs inherently hallucinate as a consequence of oral-formulaic mechanisms, the lack of authorship and ownership by LLMs may in fact serve the tech companies' interests by avoiding liability and accountability. 

Is there a way for LLMs to be a means for happiness in humanity? It seems that they will not be such a means through simpler business applications and use cases. Perhaps the only hope for human flourishing in the LLM era is for LLMs to be muses and partners in post-literate human-AI co-creation that centers attribution and meaning.


\bibliographystyle{ACM-Reference-Format}
\bibliography{singeroftales}

\end{document}